\documentclass[prl,preprint,endfloats]{revtex4-2}

\usepackage{graphicx}
\usepackage{bm}
\usepackage{pifont}
\usepackage{amsmath}
\usepackage{color}
\usepackage{comment}
\usepackage{textcomp}

\begin{document}

\title{Strain Tuning of Orbital-Driven Giant Magnetoresistance in van der Waals ferrimagnet Mn$_3$Si$_2$Te$_6$}

\author{Abdul Ahad$^{1,a}$, Miuko Tanaka$^{1}$, Shunta Aoki$^{1}$, Darius-Alexandru Deaconu$^{2}$, Varun Shah$^{2}$, Tenta Kitamura$^{3}$, Hao Ou$^{3}$, Mohammad Saeed Bahramy$^{2}$, Jiang Pu$^{3}$, Toshiya Ideue$^{1,a}$}

\affiliation{$^{1}$Institute for Solid State Physics, The University of Tokyo, Kashiwanoha, Kashiwa, Chiba 277-8581, Japan\\ 
\footnotetext[1]{Corresponding authors.
		abdulahad@issp.u-tokyo.ac.jp,
		ideue@issp.u-tokyo.ac.jp
	}
	$^{2}$Department of Physics and Astronomy, The University of Manchester, Manchester M13 9PL, United Kingdom\\
	$^{3}$Department of Physics, Institute of Science Tokyo, Tokyo 152-8552, Japan}

\begin{abstract}
Strain engineering of magnetotransport offers a powerful strategy for uncovering emergent electronic and domain phenomena in quantum magnetic materials, while providing a promising pathway toward next-generation mechanically programmable spintronic technologies. Van der Waals magnets are particularly attractive in this context because their high crystallinity and mechanical flexibility allow exceptionally large, precisely controllable strain, enabling access to strain-induced functionalities unattainable in conventional solids.
Here, we report systematic strain control of the van der Waals magnet Mn$_3$Si$_2$Te$_6$, which exhibits an unconventional colossal magnetoresistance whose microscopic origin remains under debate. We demonstrate in situ large-strain modulation of the electrical resistance in bulk crystals and show that the effect can be consistently explained by strain-tunable chiral orbital-current domains. Furthermore, measurements on exfoliated flake devices containing a single chiral domain reveal direct strain control of the electronic structure affected by orbital magnetic moment, establishing a unified microscopic mechanism for the unconventional colossal magnetoresistance. These results identify strain as an exceptionally effective control parameter for tailoring electronic and magnetic states in van der Waals magnets and provide a conceptual framework for realizing spin-straintronic functionalities based on orbital degrees of freedom.

\end{abstract}

\maketitle
\clearpage
The magnetoresistance effect serves as an extremely effective probe for quantum materials, providing valuable insights into exotic electronic states. When its magnitude is exceptionally large, it also holds potential for application in various magnetic functional devices. Examples include giant magnetoresistance in various topological materials and the colossal magnetoresistance effect in strongly correlated oxides~\cite{Shekhar2015NatPhys,Breunig2017NatCommun,Tokura2006RepProgPhys}. The unconventional giant magnetoresistance recently discovered in Mn$_3$Si$_2$Te$_6$ has also sparked intense debate regarding its mechanism due to its unique characteristics~\cite{Gu2024NatCommun,Zhang2022Nature,Seo2021Nature,Ni2021PRB}. Having Mn$^{2+}$ (3d$^5$) with S = 5/2 and L = 0, it was proposed that the spin-orbit coupling between Mn spins and orbital angular momentum coming from the in-plane Te 5$p$ orbitals induce the gap between the degenerate chiral bands and metallicity appears as one of the chiral bands crosses the Fermi level (E$_F$)~\cite{Seo2021Nature}. Crucially, this explanation requires a specific critical magnetic field for the spin to align along the $c$-axis, allowing the Fermi level to touch the band and trigger an insulator-metal transition. However, this appears to contradict the behavior observed experimentally in bulk crystals, where applying a magnetic field immediately reduces resistance~\cite{Ni2021PRB}. Therefore, a fascinating alternative mechanism has been proposed where the spontaneous formation of disordered loops of chiral orbital currents (COCs) have been assumed which will order under the magnetic field (much lower that require for out of plane spin polarization) and can show metallic behavior~\cite{Zhang2022Nature}. The COC explanation for bulk single crystals appears promising, but devices using exfoliated samples ($\sim$ 100 nm thick and several micrometers in size) exhibit different magnetoresistance behavior~\cite{Tan2024NanoLett}, further complicating the microscopic understanding. 
Related with origins of this unconventional giant magnetoresistance effect, control of the magnetoresistance of Mn$_3$Si$_2$Te$_6$ by external fields has also been investigated. For example, its controllability via applied current and doping has been proposed~\cite{Zhang2022Nature,Zhang2024NatCommun,Huang2024PRB,Das2025PRB}. For the bulk crystal, a hydrostatic pressure study has been also performed and based on the band picture (not on the COC scenario), a gap reduction between the valence and conduction band is suggested~\cite{Susilo2024NatCommun}. In this context, systematically investigating new external field responses in both bulk samples and flake devices of Mn$_3$Si$_2$Te$_6$ may be meaningful for exploring the origin of the magnetoresistance effect, which is still not fully understood, as well as for discovering new functionalities in van der Waals magnets.
Here, we demonstrate uniaxial-strain control of unconventional magnetoresistance in both bulk crystals and exfoliated flakes of Mn$_3$Si$_2$Te$_6$. Unlike hydrostatic pressure, uniaxial strain modifies crystal symmetry and anisotropy, enabling directional tuning of electronic states~\cite{Dong2023ScrMater,Dong2023NatNano,Li2025NatCommun,Mutch2019SciAdv,Dashwood2023NatCommun,Hicks2014Science,Noad2023Science,Hicks2025AnnRev}. In bulk crystals, strain effectively manipulates COC domains, allowing reversible and bidirectional resistance control through domain creation and annihilation under tensile and compressive strain. In contrast, in flake devices the COC multidomain picture no longer dominates, and transport is governed primarily by the strain-tunable band structure via modulation of spin–orbit coupling. These results establish strain as a decisive control parameter for this material, provide a unified interpretation of the contrasting transport behaviors in bulk and micro-scale flakes, and reveal a route toward strain-engineered functionalities in van der Waals magnets.

\begin{figure}
	\begin{center}
		\includegraphics*[width=13cm]{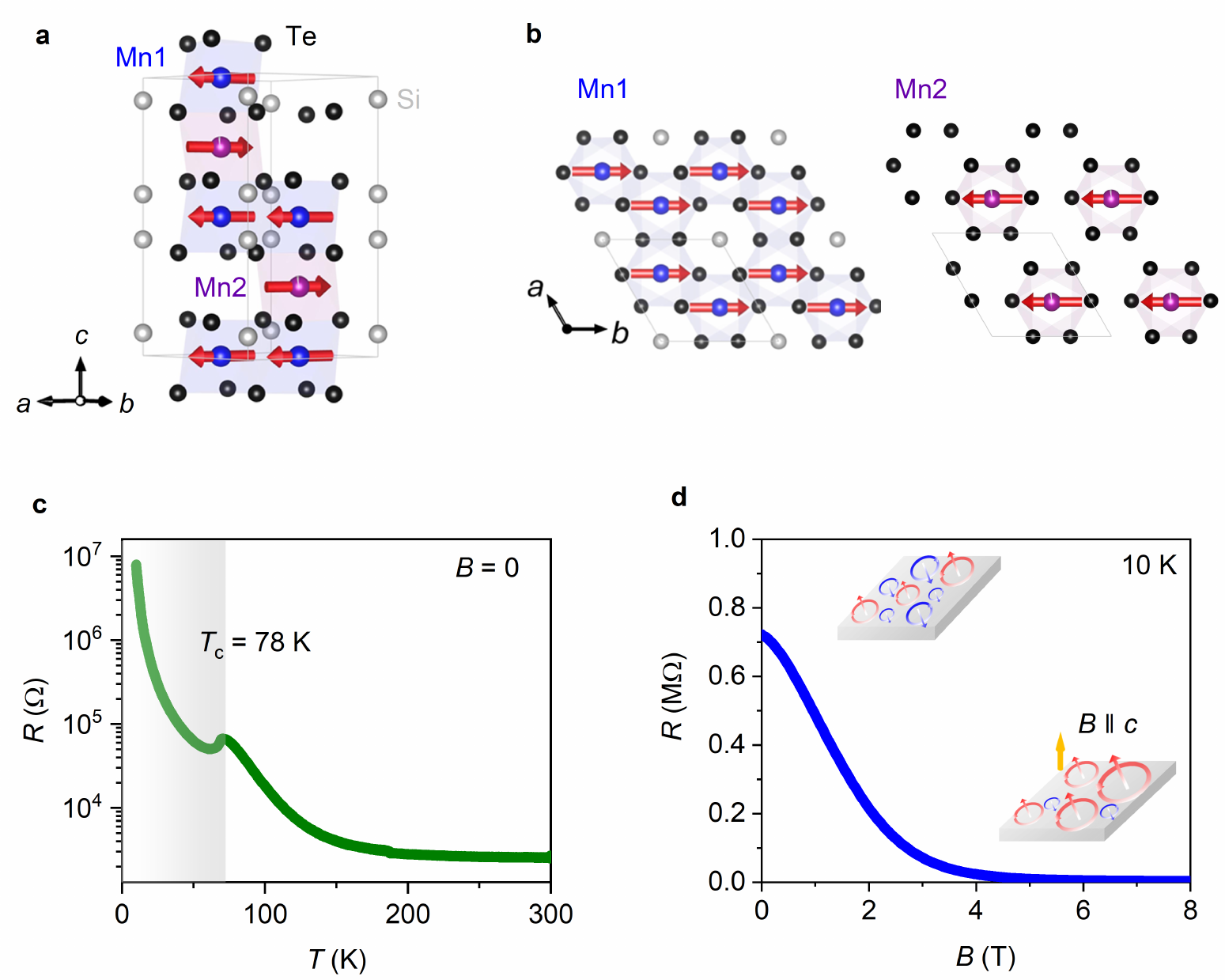}
		\caption{Side view (a) and top view (b) of the crystal/ magnetic structures. There are two magnetic sites termed as Mn1 and Mn2, in which spins are aligned ferromagnetically within a plane and couple antiferromagnetically between neighboring layers (Mn1 and Mn2). (c) Temperature dependence of the resistance at zero magnetic field (shaded area corresponds to the ferrimagnetic region). (d) Resistance as a function of magnetic field at 10 K. Inset shows the schematic of the disordered COC state (top left) at B = 0 and ordered COC state under the magnetic field.}
		\label{Fig1}
	\end{center}
\end{figure}

Mn$_3$Si$_2$Te$_6$ is a self-intercalated van der Waals magnet, in which there are two manganese sites: Mn1 and Mn2. In paramagnetic phase, space group is $P$~-31c. Below T$_c$ = 78 K, both sites order ferromagnetically within the same layer, while the interlayer ordering is antiferromagnetic~\cite{May2017PRB} (see Figure~\ref{Fig1}a). The neutron diffraction study suggested strong easy-plane anisotropy but making $\sim$ 10$^{\circ}$ angle with the $ab$-plane (see supplementary material for the magnetization measurements for the anisotropy related discussion). This arrangement can be explained under the monoclinic symmetry~\cite{Lovesey2023PRB,Ye2022PRB} of the magnetic space group $C$~2'/c'. Such magnetic symmetry allows ferromagnetism and the piezomagnetic effect, where bi-directional strain control is expected if magnetic order is related to a physical property (e.g., electric transport)~\cite{Liu2019AdvElectronMater,Guo2020AdvMater}. The Mn1 layer has hexagonal symmetry while Mn2 has trigonal arrangement (see Figure~\ref{Fig1}b). The tellurium atoms make a hexagonal plane where the domains of chiral orbital current (COC) loops can appear~\cite{Zhang2022Nature}. There have been several reports that such chiral orbital currents (or loop current order) can play the role in novel quantum materials and their resultant exotic physical properties such as spin nematic phase~\cite{Bounoua2020CommunPhys}, fluctuating loop induced superconductivity~\cite{Palle2024SciAdv}, charge ordering~\cite{Mielke2022Nature}, etc. In the case of Mn$_3$Si$_2$Te$_6$, it is proposed that the loop current is of Varma type, which was originally discussed for the cuprates where the phase associated with the hopping parameter between nearest neighbor can emerge as an orbital current~\cite{Varma1997PRB,Bulut2015PRB}. Although there is no direct evidence of COC domains in Mn$_3$Si$_2$Te$_6$ in the literature, a recent spin resonance study demonstrated the existence of finite orbital angular momentum L$_z$\cite{Tanaka2025PRB}. Moreover, as discussed in Ref~\cite{Zhang2022Nature}, we also did not observe features supporting the additional impurity band model~\cite{Seo2021Nature} (see supplementary Figure S2), which suggests that the COC picture is more suitable here. 
Figure~\ref{Fig1}c represents the zero-field resistance measurement as a function of temperature, featuring a ferrimagnetic ordering temperature around T$_c$ = 78 K. It is worth noting here that resistance increases while lowering the temperature and the disordered domains of COC are responsible for such enhancement (the band gap picture is not applicable in the bulk sample here as discussed in the following). Figure ~\ref{Fig1}d displays the magnetoresistance spanning from zero field where the spontaneous disordered loops persist, to the applied field of 8 T (along the $c$ direction) which makes the domain of one chirality grow, while the other is suppressed (schematically shown in the insets). This ordered COC domain state is responsible for the unconventional CMR state and could be tunable by other external perturbations as it is dependent on the nearest neighbors and the spatial symmetry.

\begin{figure}
	\begin{center}
		\includegraphics*[width=13cm]{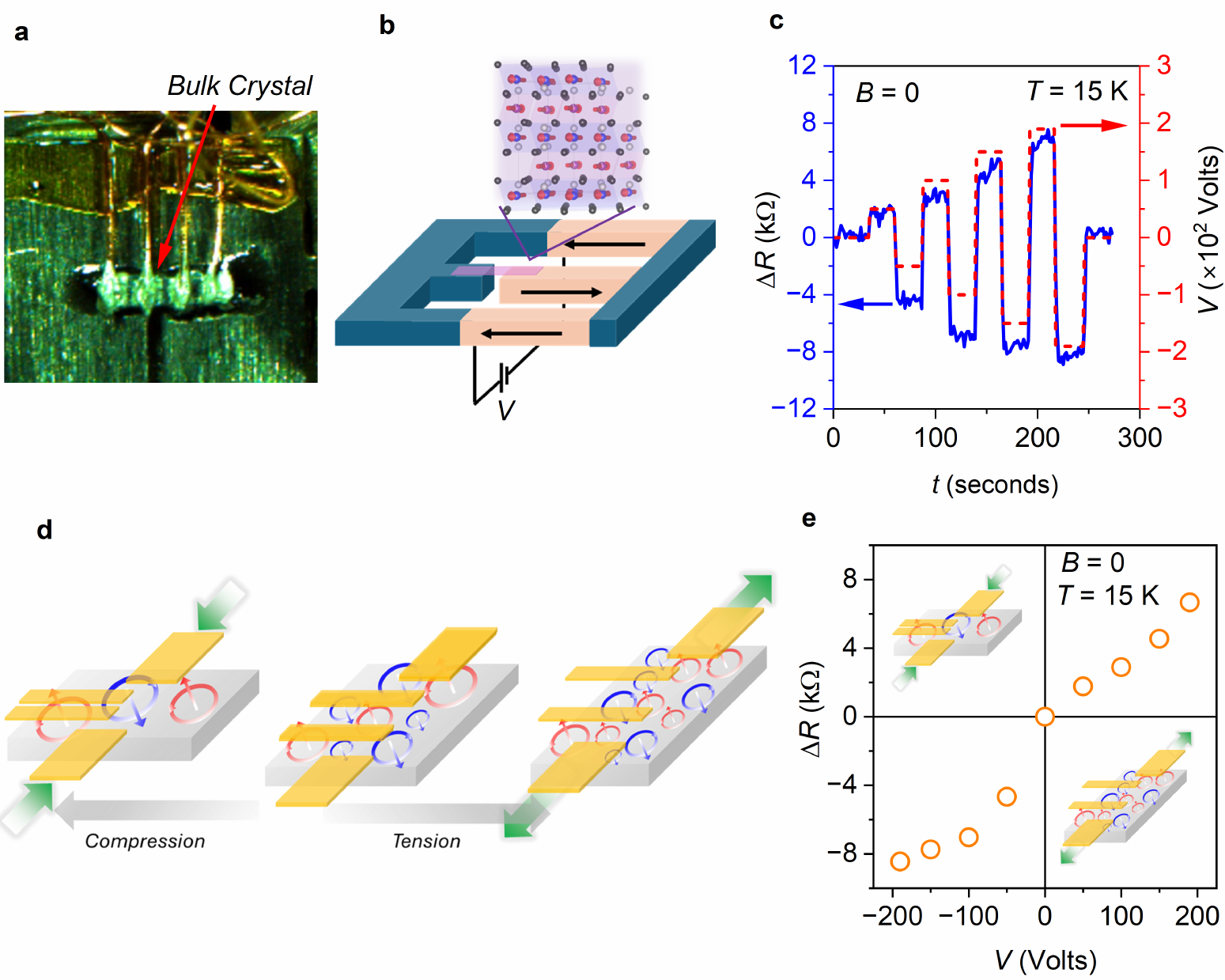}
		\caption{(a) Photograph of a bulk crystal mounted on the holder of strain cell’s titanium plate trench. (b) Schematic of strain cell. It contains two parts: rigid arms and free arms. Sample is mounted, bridging both arms where applied voltage can control the movement of free side.  (c) Resistance modulation ($\Delta R =R(V \neq 0)-R(V=0)$) under the strain (application period of $\sim$ 30 seconds) of tensile and compressive nature, recorded at 15 K in zero magnetic field. (d) Schematic of the COC domains under tensile (compressive) strain. Left schematic corresponds to compressive case where fewer domains exist, Middle is the unstrained sample while the Right schematic corresponds to the creation of smaller domains/ breaking of bigger into smaller domains. (e) Resistance modulation ($\Delta$ R) vs. applied voltages for compressive and tensile strain at 15 K.}
		\label{Fig2}
	\end{center}
\end{figure}

As discussed above, the COC state is disordered at B = 0 and consists of both domains (Right and left moving). Unlike hydrostatic pressure, uniaxial strain can break rotational symmetries~\cite{Chu2012Science,Du2021NatRevPhys}. Consequently, such a strain could be utilized to manipulate the domains (which are protected by rotational symmetry). In the current study, we applied the uniaxial strain on the single crystal of Mn$_3$Si$_2$Te$_6$. We employed strain cell (Razorbill instruments) to apply the strain (see Method for details). Figure~\ref{Fig2}a shows the photograph of the crystal mounted on the strain cell holder via epoxy. A thin bulk crystal (thickness of $\sim$100 $\mu$m) is used to apply the large strain. The strain cell uses piezo-stacks which stretches or compresses via the applied voltages polarities~\cite{Hicks2014Science}. The schematic shown in Figure~\ref{Fig2}b, where two arms of the cell stretch or compress depend on the polarity of the applied voltage, i.e., positive for the tensile strain and negative for the compressive strain. The exact value of applied strain is not estimated here but remains proportional to the voltages applied. Figure~\ref{Fig2}c displays the in-situ resistance modulation ($\Delta R =R(V \neq 0)-R(V=0)$) with the stretch/compress uniaxial strain, recorded at 15 K in the zero magnetic field. From the results, the positive (negative) bias of piezo stack i.e., tensile (compressive) strain, change the sample resistance dramatically ($\Delta R \sim k \Omega)$, causing the higher (lower) resistance state. The effect of uniaxial strain is linear in nature (see Figure~\ref{Fig2}e) and could be linked to the piezomagnetic effect, which linearly couples the magnetization ($M$) to the strain ($\epsilon$) (i.e., $M= \Lambda \epsilon$, where $\Lambda$ is the piezomagnetic tensor). This offers a great deal of bidirectional functionality and could be useful in zero field mechanically programmable devices~\cite{Ikhlas2022NatPhys}. 
The modulation of resistance under strain in the bulk single crystal can be understood in terms of a strain-driven reorganization of chiral orbital current (COC) domains. Under tensile strain, the anisotropic elastic response favors subdivision of large domains into multiple smaller domains, leading to an increased density of domain walls along the current path, as schematically illustrated in Figure~\ref{Fig2}d. This enhanced domain-wall density results in a higher resistive state. Conversely, under compressive strain, neighboring domains tend to coalesce, reducing the number of domain walls intersecting the current flow and thereby lowering the resistance. Uniaxial strain modifies the electronic hopping amplitudes anisotropically. Since the energy of a COC domain boundary is controlled by the cost of distorting the orbital-current pattern across neighboring bonds, a reduction of hopping along the strain axis lowers the energy penalty for domain walls whose normal lies along that direction. Consequently, strain introduces a directional preference for domain-wall formation. Notably, this substantial strain-induced change in resistance is achieved in the absence of any applied magnetic field. Below, we address the realization of a single-domain COC state by reducing the lateral dimensions of the flake.

\begin{figure}
	\begin{center}
		\includegraphics*[width=13cm]{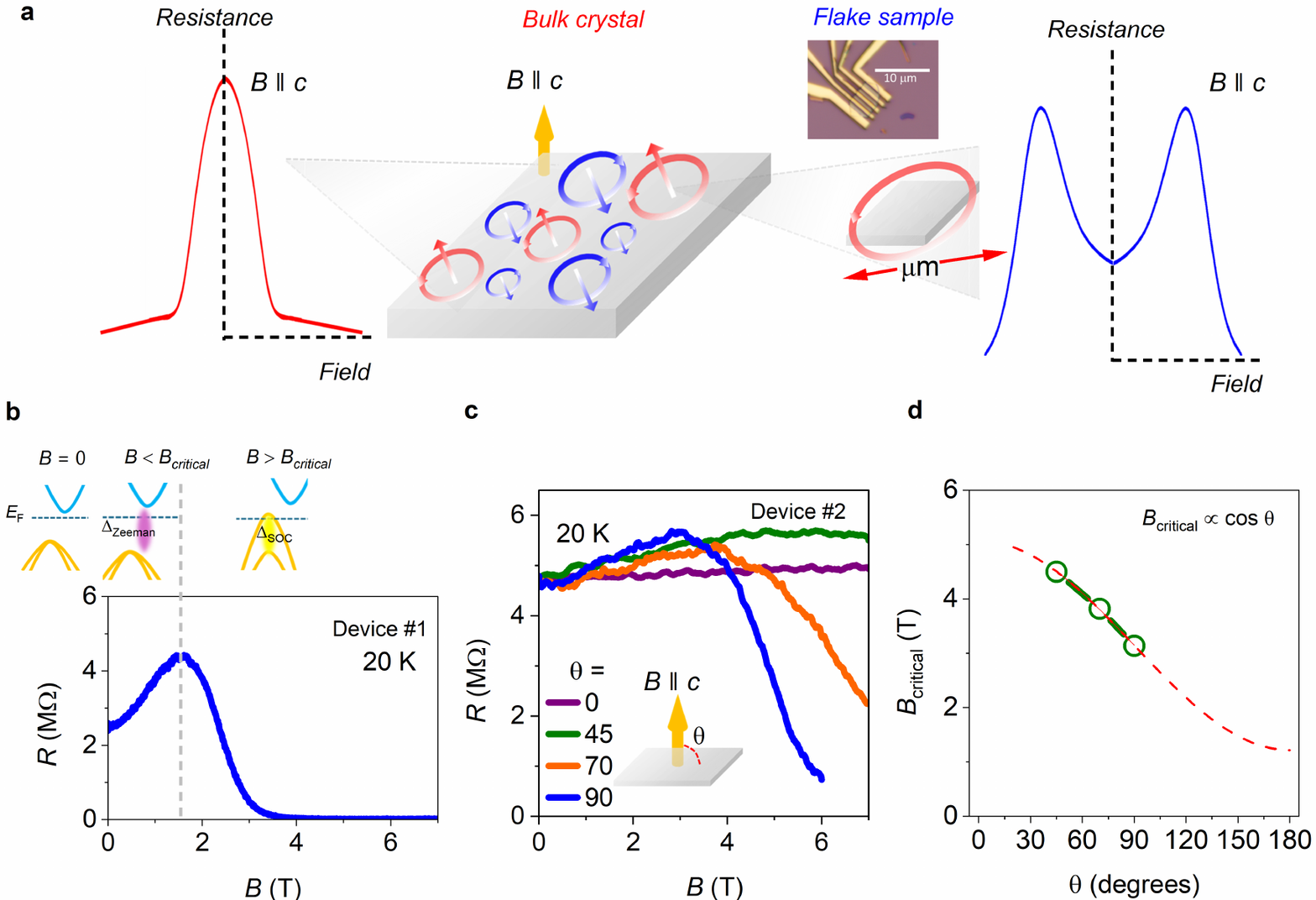}
		\caption{(a) Schematics of magnetoresistance (B $\parallel$ c) of bulk samples containing multidomain COC loops (Left panel) and that of micrometer size flake sample with single domain (Right panel). Optical microscope image of one of the representative devices is also shown. A peak structure appears at the intermediate field in the magnetoresistance of flake samples.  (b) Magnetoresistance at T = 20 K for the exfoliated flake (device $\#$1). Three regions can be seen: B = 0, B $<$ B$_{critical}$ and B $>$ B$_{critical}$ (see text for details). (c) Magnetoresistance peak marked as “B$_{critical}$” shifts towards the higher field values with changing the angle $\theta$ from 90$^{\circ}$ (B $\parallel$ c) to 0$^{\circ}$ (B $\parallel$ $ab$). B$_{critical}$ is linked with the onset of the SOC dominant region and hence with the L$_z$. The B$_{critical}$ trend with the crystal axis angle is plotted in (d). Dashed red line corresponds to the cosine function fitted over dataset.}
		\label{Fig3}
	\end{center}
\end{figure}

In bulk crystals, it is expected that multidomain state of COC governs the magnetotransport as explained earlier (Figure~\ref{Fig3}a left panel). In such samples, the multi-domain state hinders the intrinsic band picture of Mn$_3$Si$_2$Te$_6$. In contrast, we propose that for a micrometer-scale nanoflake, only a single domain can exist (as shown by single red loop; size of the COC loop $\sim$ few $\mu$m)~\cite{Gu2024NatCommun}, the magnetoresistance shows totally different behavior~\cite{Tan2024NanoLett} (Figure~\ref{Fig3}a right panel). Figure~\ref{Fig3}b shows the magnetoresistance recorded for the nanoflake (Device $\#$1) obtained from the same batch of the bulk crystals. It highlights three different regions: B = 0, B $<$ B$_{critical}$ and B $>$ B$_{critical}$. At B=0 the system is consistent with a gapped state, corresponding to a high resistance regime. With increasing magnetic field (B $\parallel$ c), the resistance increases. This could be understood as a Zeeman-driven gap enhancement. Beyond a finite critical field (B$_{critical}$), an alternative regime emerges. We speculate that spin-orbit coupling (L $\parallel$ S) lifts band degeneracies, leading to band crossings near the Fermi energy and the onset of metallic behavior. All the devices measured (lateral size of $\sim$ 7 x 4 $\mu$m) show similar behavior (see supplementary section 7). The details of this expectation of band evolution are discussed in the supplemental material section 9. Note that out of plane orbital angular momentum is crucial for this mechanism. In Figure~\ref{Fig3}c we have recorded the magnetoresistance as a function of the out of plane magnetic field by rotating the sample from angle 90$^{\circ}$ (B $\parallel$ $c$ axis) to 0$^{\circ}$ (B $\parallel$ $ab$ plane). It was found that the peak position in MR marked as B$_{critical}$ depends on the field direction; i.e., B$_{critical}$ moves towards the higher field as the angle $\theta$ increases (see Figure~\ref{Fig3}d). In this plot, the dashed line corresponds to the tendency B$_{critical}$ $\propto$ cos $\theta$. This observation clearly explains the crucial role of the out-of-plane component of L$_z$ and hence confirms the intrinsic picture of bands involved in the magnetoresistance of flake Mn$_3$Si$_2$Te$_6$ without invoking multidomains of COC or incorporating additional impurity bands to close the gap at the low magnetic field~\cite{Zhang2022Nature,Seo2021Nature}. These observations provide unambiguous evidence of unique differences between bulk and flake samples of Mn$_3$Si$_2$Te$_6$. In the following we will investigate the tunability of the single domain COC state in flake samples.

\begin{figure}
	\begin{center}
		\includegraphics*[width=13cm]{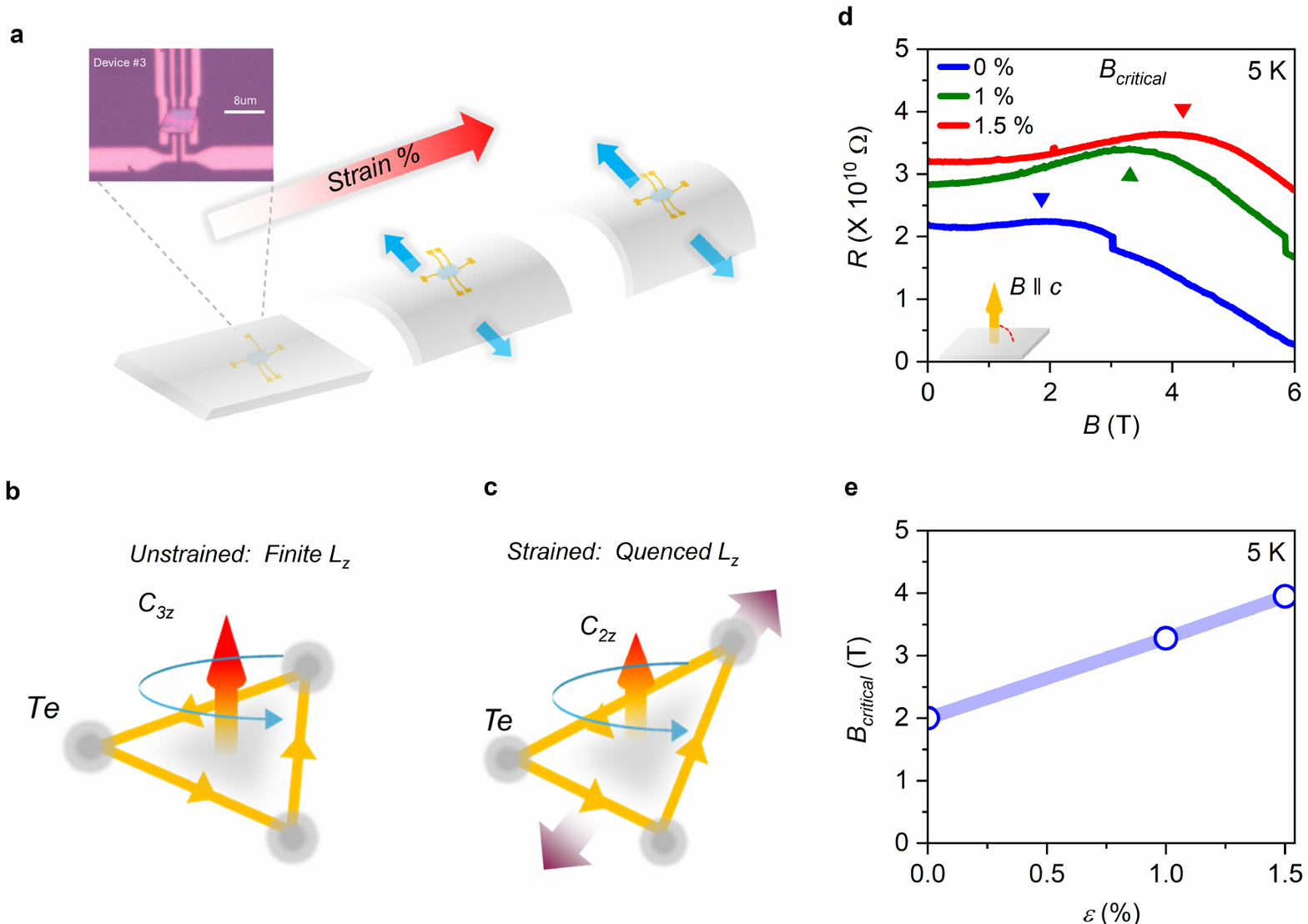}
		\caption{(a) Schematic of uniaxial strain application for the nanodevices fabricated on the flexible substrate. Inset shows the optical microscope image of the device used in the measurements. (b), (c) Schematics of the orbital angular momentum originating from Te sites in unstrained samples with $C_{3z}$ symmetry (b) and its quenching due to the symmetry reduction under uniaxial strain (c). (d) Magnetoresistance at T = 5 K under 0 $\%$, 1 $\%$ and 1.5 $\%$ tensile strain. With increasing the strain, Bcritical shifted towards higher magnetic field (e).}
		\label{Fig4}
	\end{center}
\end{figure}

As discussed earlier, the single COC domain state can be selected by reducing the lateral dimensions of the Mn$_3$Si$_2$Te$_6$ sample. Flake samples exfoliated on the flexible substrate were used to achieve single COC domain states and manipulate the intrinsic band structure by large uniaxial strain. Schematic (Figure~\ref{Fig4}a) comprises the technique used~\cite{Ahad2026arXiv} for the tensile strain application on the nano devices fabricated on the flexible substrate (see Methods and supplementary for the details). The flat device corresponds to the 0 $\%$ strain case while the curved devices are imposing uniaxial tensile strain (strain direction is shown by blue arrows) along the straight edge of the flake which we chose as a directional reference for the multiple devices. The orbital angular momentum (L$_z$) of Mn$_3$Si$_2$Te$_6$ originates from the Te $p_x$, $p_y$ orbitals in the $ab$ plane; The Te trimer (Fig.~\ref{Fig4}b) has rotational symmetry $C_{3z}$ and thus gives a finite phase $\sim$ $e^{(\pm i \theta)}$ during the electron hopping and resultant out of plane L$_z$~\cite{Seo2021Nature}. 
Figure~\ref{Fig4}d shows the magnetoresistance at T = 5 K under 0 $\%$ (blue), 1 $\%$ (green) and 1.5 $\%$ (red) tensile strain values. Systematic modulation of the magnetoresistance by uniaxial strain has been observed in the flake sample with a single COC domain. Importantly, Bcritical linearly increases with the uniaxial strain value (Fig.~\ref{Fig4}c), implying the L$_z$ reduction by the uniaxial strain application. In the loop-current scenario, the orbital moment is proportional to the complex hopping amplitudes coming from the frustration of the electronic hopping~\cite{Bounoua2020CommunPhys}. Application of the uniaxial strain in the $ab$ plane will stretch the lattice, switching the system from $C_{3z}$ to lower $C_{2z}$ symmetry. It is expected that this symmetry change renormalizes the hopping parameters and the effective plaquette flux, thus quenching the loop-induced orbital angular momentum. Both observation of the magnetic field directional dependence of B$_{critical}$ in the unstrained sample (B$_{critical}$ $\propto$ cos $\theta$) and its uniaxial strain dependence (B$_{critical}$ $\propto$ $\epsilon$) cumulatively support our claim of the role of L$_z$ for the crossover peak behavior in the magnetoresistance.

In conclusion, we have demonstrated the uniaxial strain tunability of resistance in the van der Waals ferrimagnet Mn$_3$Si$_2$Te$_6$. In both the bulk crystal and the nanoflake the resistance can be largely modulated but via different mechanisms. In bulk crystals, COC domains can be manipulated via tensile/compressive strain. On the other hand, in flake samples with single COC domains, the band picture becomes valid, where strain can tune the magnitude of orbital angular momentum and spin orbital coupling. Our results not only demonstrate the controllability of exotic electronic states in van der Waals magnets under mechanical perturbation but also solve the long-standing issue of differing magnetotransport behavior between bulk and micro size crystals.

\section*{Data availability}

The data presented in the current study are available from the corresponding authors on reasonable request.

\section*{Acknowledgements} T.I. was supported by Japan Society for the Promotion of Science (JSPS) KAKENHI (Grant Numbers JP23H00088, JP24H01176, JP25H00839, JP25H02117), JST FOREST (Grant Number JPMJFR213A), and Murata Science and Education Foundation (Grant Number M24AN111). N.D.K was supported by JSPS KAKENHI Grant Numbers JP25K00956 and JP23K13069. J.P. was supported by JSPS KAKENHI (Grant Numbers 21H05232, 21H05236) and JST FOREST (Grant Number JPMJFR223Z). A.A. acknowledges support from JSPS KAKENHI (Grant Number 23KF0195).

\bibliography{Main.bib}

@article{Shekhar2015NatPhys,
  author = {Shekhar, C. and others},
  title = {Extremely large magnetoresistance and ultrahigh mobility in the topological Weyl semimetal candidate NbP},
  journal = {Nature Physics},
  volume = {11},
  pages = {645--649},
  year = {2015}
}

@article{Breunig2017NatCommun,
  author = {Breunig, O. and others},
  title = {Gigantic negative magnetoresistance in the bulk of a disordered topological insulator},
  journal = {Nature Communications},
  volume = {8},
  pages = {15545},
  year = {2017}
}

@article{Tokura2006RepProgPhys,
  author = {Tokura, Y.},
  title = {Critical features of colossal magnetoresistive manganites},
  journal = {Reports on Progress in Physics},
  volume = {69},
  pages = {797--851},
  year = {2006}
}

@article{Gu2024NatCommun,
  author = {Gu, Y. and others},
  title = {Unconventional insulator-to-metal phase transition in Mn3Si2Te6},
  journal = {Nature Communications},
  volume = {15},
  year = {2024}
}

@article{Zhang2022Nature,
  author = {Zhang, Y. and others},
  title = {Control of chiral orbital currents in a colossal magnetoresistance material},
  journal = {Nature},
  volume = {611},
  pages = {467--472},
  year = {2022}
}

@article{Seo2021Nature,
  author = {Seo, J. and others},
  title = {Colossal angular magnetoresistance in ferrimagnetic nodal-line semiconductors},
  journal = {Nature},
  volume = {599},
  pages = {576--581},
  year = {2021}
}

@article{Ni2021PRB,
  author = {Ni, Y. and others},
  title = {Colossal magnetoresistance via avoiding fully polarized magnetization in the ferrimagnetic insulator Mn3Si2Te6},
  journal = {Physical Review B},
  volume = {103},
  year = {2021}
}

@article{Tan2024NanoLett,
  author = {Tan, C. and others},
  title = {Electrically Tunable, Rapid Spin-Orbit Torque Induced Modulation of Colossal Magnetoresistance in Mn3Si2Te6 Nanoflakes},
  journal = {Nano Letters},
  volume = {24},
  pages = {4158--4164},
  year = {2024}
}

@article{Zhang2024NatCommun,
  author = {Zhang, Y. and others},
  title = {Current-sensitive Hall effect in a chiral-orbital-current state},
  journal = {Nature Communications},
  volume = {15},
  year = {2024}
}

@article{Huang2024PRB,
  author = {Huang, C. and others},
  title = {Tuning the colossal magnetoresistance in (Mn1-xMgx)3Si2Te6 by engineering the gap and magnetic properties via doping and pressure},
  journal = {Physical Review B},
  volume = {109},
  pages = {205145},
  year = {2024}
}

@article{Das2025PRB,
  author = {Das, A. and Mukhopadhyay, S.},
  title = {Tuning the chiral orbital currents in a colossal magnetoresistive nodal-line ferrimagnet},
  journal = {Physical Review B},
  volume = {111},
  pages = {174419},
  year = {2025}
}

@article{Susilo2024NatCommun,
  author = {Susilo, R. A. and others},
  title = {High-temperature concomitant metal-insulator and spin-reorientation transitions in a compressed nodal-line ferrimagnet Mn3Si2Te6},
  journal = {Nature Communications},
  volume = {15},
  year = {2024}
}

@article{Dong2023ScrMater,
  author = {Dong, S. and others},
  title = {Strain-tuning Bloch- and Neel-type magnetic skyrmions: A phase-field simulation},
  journal = {Scripta Materialia},
  volume = {222},
  year = {2023}
}

@article{Dong2023NatNano,
  author = {Dong, Y. and others},
  title = {Giant bulk piezophotovoltaic effect in 3R-MoS2},
  journal = {Nature Nanotechnology},
  volume = {18},
  pages = {36--41},
  year = {2023}
}

@article{Li2025NatCommun,
  author = {Li, J. and others},
  title = {The classical-to-quantum crossover in the strain-induced ferroelectric transition in SrTiO3 membranes},
  journal = {Nature Communications},
  volume = {16},
  pages = {4445},
  year = {2025}
}

@article{Mutch2019SciAdv,
  author = {Mutch, J. and others},
  title = {Evidence for a strain-tuned topological phase transition in ZrTe5},
  journal = {Science Advances},
  volume = {5},
  year = {2019}
}

@article{Dashwood2023NatCommun,
  author = {Dashwood, C. D. and others},
  title = {Strain control of a bandwidth-driven spin reorientation in Ca3Ru2O7},
  journal = {Nature Communications},
  volume = {14},
  pages = {6197},
  year = {2023}
}

@article{Hicks2014Science,
  author = {Hicks, C. W. and others},
  title = {Strong Increase of Tc of Sr2RuO4 Under Both Tensile and Compressive Strain},
  journal = {Science},
  volume = {344},
  pages = {283--285},
  year = {2014}
}

@article{Noad2023Science,
  author = {Noad, H. M. L. and others},
  title = {Giant lattice softening at a Lifshitz transition in Sr2RuO4},
  journal = {Science},
  volume = {382},
  pages = {447--450},
  year = {2023}
}

@article{Hicks2025AnnRev,
  author = {Hicks, C. W. and others},
  title = {Probing Quantum Materials with Uniaxial Stress},
  journal = {Annual Review of Condensed Matter Physics},
  volume = {16},
  pages = {417--442},
  year = {2025}
}

@article{May2017PRB,
  author = {May, A. F. and others},
  title = {Magnetic order and interactions in ferrimagnetic Mn3Si2Te6},
  journal = {Physical Review B},
  volume = {95},
  year = {2017}
}

@article{Lovesey2023PRB,
  author = {Lovesey, S. W.},
  title = {Anapole, chiral, and orbital states in Mn3Si2Te6},
  journal = {Physical Review B},
  volume = {107},
  year = {2023}
}

@article{Ye2022PRB,
  author = {Ye, F. and others},
  title = {Magnetic structure and spin fluctuations in the colossal magnetoresistance ferrimagnet Mn3Si2Te6},
  journal = {Physical Review B},
  volume = {106},
  year = {2022}
}

@article{Liu2019AdvElectronMater,
  author = {Liu, Z. and others},
  title = {Antiferromagnetic Piezospintronics},
  journal = {Advanced Electronic Materials},
  volume = {5},
  year = {2019}
}

@article{Guo2020AdvMater,
  author = {Guo, H. and others},
  title = {Giant Piezospintronic Effect in a Noncollinear Antiferromagnetic Metal},
  journal = {Advanced Materials},
  volume = {32},
  year = {2020}
}

@article{Bounoua2020CommunPhys,
  author = {Bounoua, D. and others},
  title = {Loop currents in two-leg ladder cuprates},
  journal = {Communications Physics},
  volume = {3},
  pages = {123},
  year = {2020}
}

@article{Palle2024SciAdv,
  author = {Palle, G. and others},
  title = {Superconductivity due to fluctuating loop currents},
  journal = {Science Advances},
  volume = {10},
  year = {2024}
}

@article{Mielke2022Nature,
  author = {Mielke, C. and others},
  title = {Time-reversal symmetry-breaking charge order in a kagome superconductor},
  journal = {Nature},
  volume = {602},
  pages = {245--250},
  year = {2022}
}

@article{Varma1997PRB,
  author = {Varma, C. M.},
  title = {Non-Fermi-liquid states and pairing instability of a general model of copper oxide metals},
  journal = {Physical Review B},
  volume = {55},
  pages = {14554--14580},
  year = {1997}
}

@article{Bulut2015PRB,
  author = {Bulut, S. and Kampf, A. P. and Atkinson, W. A.},
  title = {Instability towards staggered loop currents in the three-orbital model for cuprate superconductors},
  journal = {Physical Review B},
  volume = {92},
  pages = {195140},
  year = {2015}
}

@article{Tanaka2025PRB,
  author = {Tanaka, M. and others},
  title = {Magnetic resonance and microwave resistance modulation in the van der Waals ferrimagnet Mn3Si2Te6},
  journal = {Physical Review B},
  volume = {112},
  pages = {L180401},
  year = {2025}
}

@article{Chu2012Science,
  author = {Chu, J.-H. and others},
  title = {Divergent Nematic Susceptibility in an Iron Arsenide Superconductor},
  journal = {Science},
  volume = {337},
  pages = {710--712},
  year = {2012}
}

@article{Du2021NatRevPhys,
  author = {Du, L. and others},
  title = {Engineering symmetry breaking in 2D layered materials},
  journal = {Nature Reviews Physics},
  volume = {3},
  pages = {193--206},
  year = {2021}
}

@article{Ikhlas2022NatPhys,
  author = {Ikhlas, M. and others},
  title = {Piezomagnetic switching of the anomalous Hall effect in an antiferromagnet at room temperature},
  journal = {Nature Physics},
  volume = {18},
  pages = {1086--1093},
  year = {2022}
}

@article{Ahad2026arXiv,
  author = {Ahad, A. and others},
  title = {Piezomagnetic transport in van der Waals noncoplanar Antiferromagnets},
  journal = {arXiv},
  year = {2026},
  eprint = {2602.04245}
}

\clearpage

\end{document}